\documentclass{iau}
\usepackage{graphicx}
\usepackage{rotating}
\usepackage{natbib}
\usepackage{txfonts}
\usepackage{url}
\def\itSigma{{\mathit\Sigma}}
\def\SigmaUnit{W m$^{-2}$\,Hz$^{-1}$\,sr$^{-1}$}
\usepackage[svgnames]{xcolor}
\makeatletter
\def\ps@plain{\let\@mkboth\@gobbletwo
     \def\@oddhead{\@journal}
     \let\@evenhead\@empty
     \let\@evenhead\@journal
     \def\@oddfoot{\reset@font{\vbox to 0pt{\vskip 50pt
     \parbox{13cm}{\raggedright\textcolor{DarkGreen}{\footnotesize\bf To
     appear in: ``Supernova environmental impacts'', eds A.\ Ray \&
     R.\ McCray, IAU Symposium No. 296 (Cambridge University Press),
     in press, 2013.}}\vss}}\hskip-7cm\thepage\hss}
     \let\@evenfoot\@oddfoot}
\def\@oddfoot{\vbox to 0pt{\vskip 50pt
\parbox{13cm}{\raggedright\textcolor{DarkGreen}{\footnotesize\bf To
appear in: ``Supernova environmental impacts'', eds A.\ Ray \&
R.\ McCray, IAU Symposium No. 296 (Cambridge University Press),
in press, 2013.}}\vss}}
\let\@evenfoot=\@oddfoot
\makeatother
\title{The Galactic distribution of SNRs}

\author{D.~A.~Green}

\affiliation{Cavendish Laboratory, 19 J.~J.~Thomson Ave.,
             Cambridge, CB3 0HE, U.K.\\
email: {\tt dag@mrao.cam.ac.uk}}

\pubyear{2013}
\volume{296}
\pagerange{\pageref{firstpage}-\pageref{lastpate}}
\jname{Supernova environmental impacts}
\editors{A.\ Ray \& R.\ McCray eds}
\begin{document}
\maketitle

\label{firstpage}

\begin{abstract}
It is not straightforward to determine the distribution of supernova
remnants (SNRs) in the Galaxy. The two main difficulties are that there
are observational selection effects that mean that catalogues of SNRs
are incomplete, and distances are not available for most remnants. Here
I discuss the selection effects that apply to the latest catalogue of
Galactic SNRs. I then compare the observed distribution of `bright' SNRs
in Galactic longitude with that expected from models in order to
constrain the Galactic distribution of SNRs.
\end{abstract}

\firstsection

\section{Introduction}

The distribution of supernova remnants (SNRs) within the Galaxy is of
interest for a variety of reasons, not least because they are important
sources of energy and high energy particles in the Galaxy. I discuss
here the observational selection effects that make current catalogues of
SNRs incomplete, and the difficulties in obtaining distances for most
remnants. Both of these issues make it difficult to derive the Galactic
distribution of SNRs directly. I present constraints on the distribution
of SNRs with Galactocentric radius, by comparison of the distribution of
bright remnants with Galactic longitude with those expected from simple
models. These results are similar to those presented in
\cite{2012AIPC.1505....5G}, but here I concentrate more on a discussion
of the selection effects that apply to current SNR catalogues. In
addition, the analysis presented here excludes the region near
$l=0^\circ$, where the observational selection effects are extreme.

\section{Background}

I have produced several catalogues of Galactic SNRs. The earliest
version, from 1984, contained 145 remnants \citet{1984MNRAS.209..449G}.
The number of known remnants has almost doubled in the following 25
years, with the most recent version \citet{2009BASI...37...45G}
containing 274 SNRs. Note, however, that there are many other possible
and probable remnants that have also been proposed, which are briefly
described in the documentation for the web version of the
catalogue\footnote{See: \url{http://www.mrao.cam.ac.uk/surveys/snrs/}}.
These objects are not included in the main catalogue of 274 remnants, as
further observations are required to confirm their nature, or their
parameters, e.g.\ their full extent. The largest increases in the number
of identified remnants are due to large area Galactic radio surveys,
e.g.\ the Effelsberg 2.7-GHz survey and the MOST survey, see
Section~\ref{s:selection}.

There are two problems that make it difficult to derive the Galactic
distribution of SNRs directly: (i) there are significant observational
selection effects that means that the catalogue of SNRs in incomplete,
and (ii) distances are not available for all SNRs. These two issues are
discussed further in the next two subsections.

\begin{figure}
\centerline{\includegraphics[angle=270,width=8.5cm]{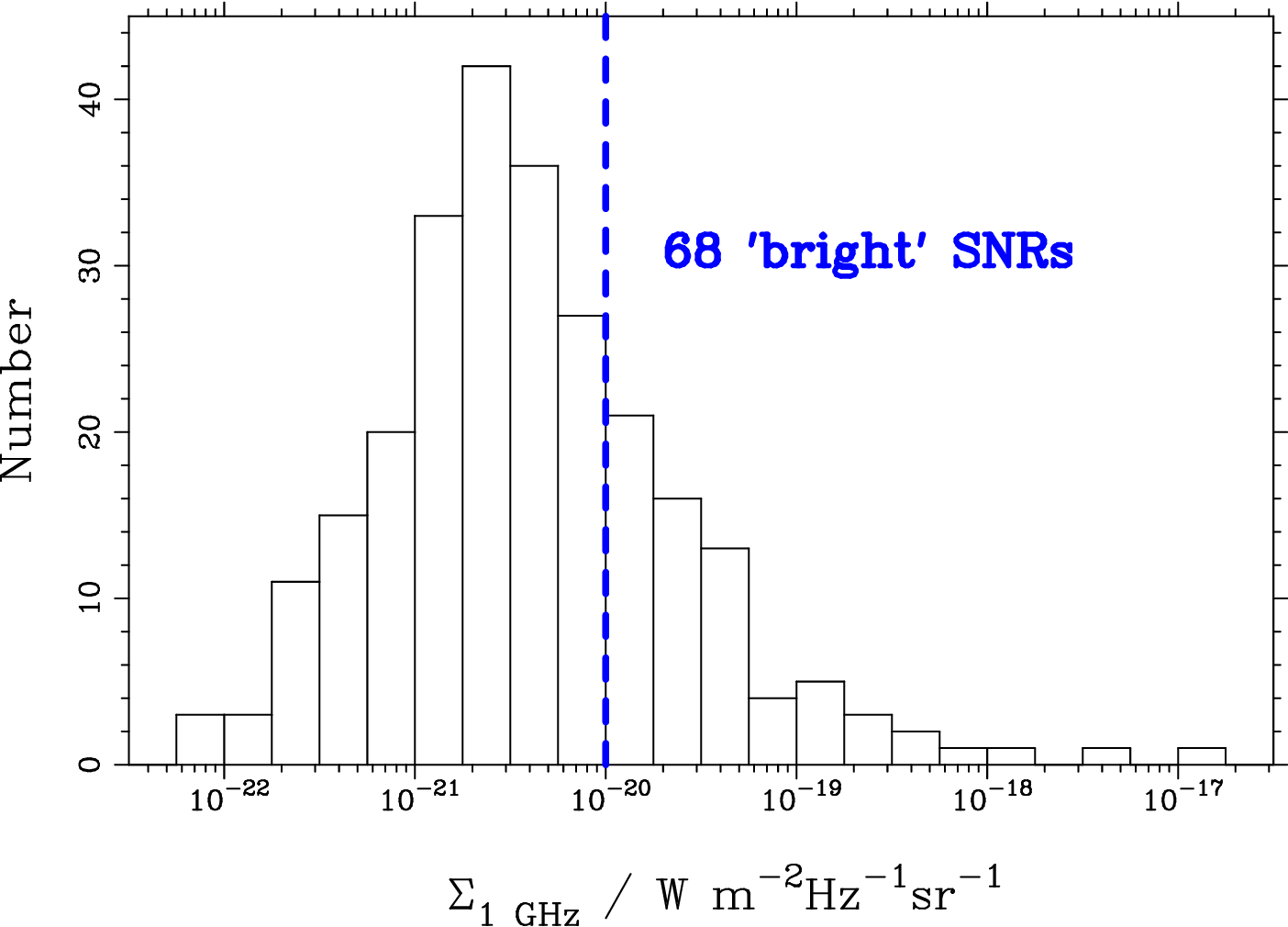}}
\bigskip
\centerline{\includegraphics[angle=270,width=8.5cm]{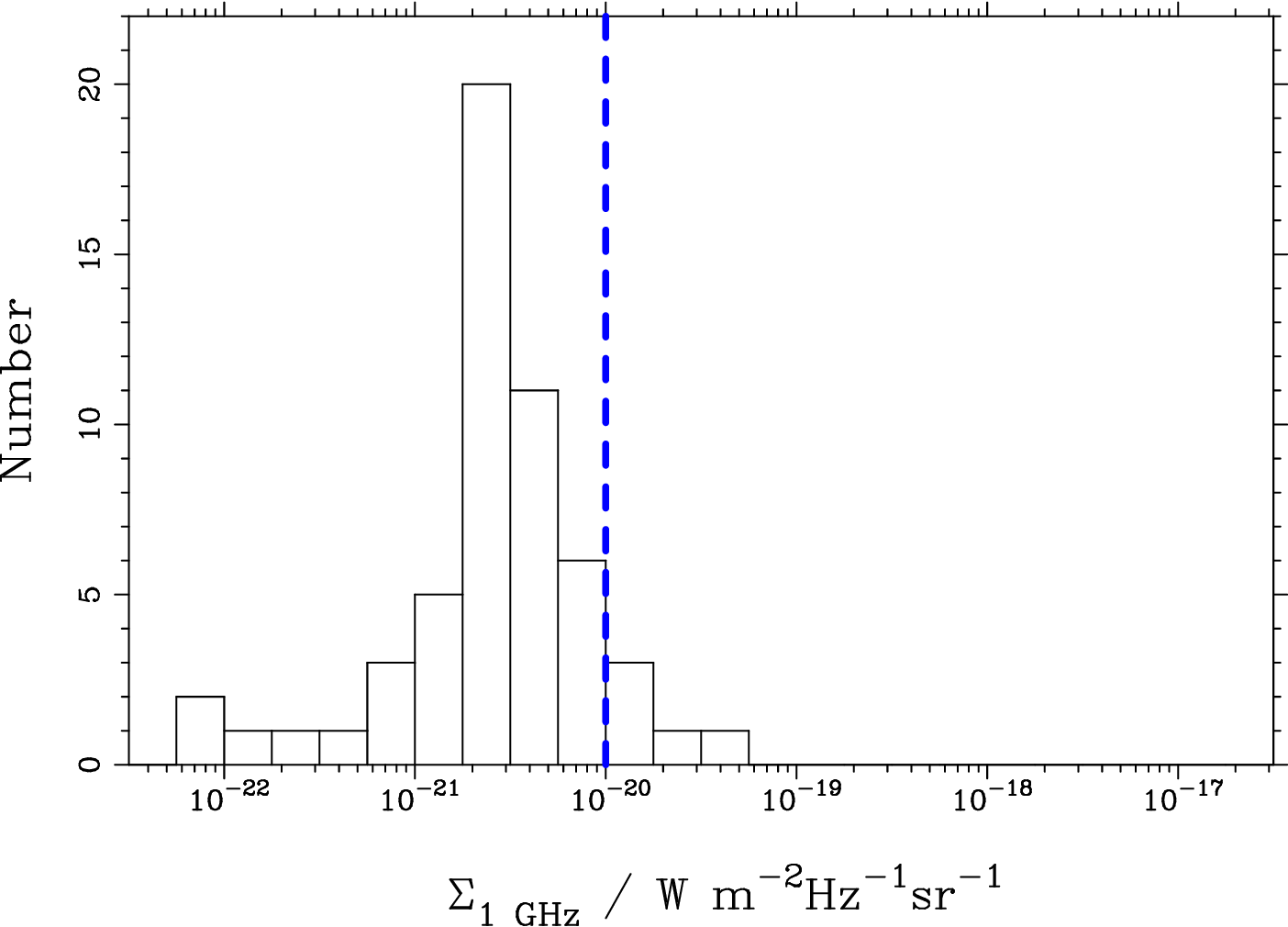}}
\bigskip
\caption{Histograms of the 1-GHz surface brightness of: (top) all
catalogued SNRs, and (bottom) those in the area covered by the
Effelsberg 2.7-GHz survey added to the catalogue since
1991.}\label{f:sigma}
\end{figure}

\begin{figure}
\centerline{\includegraphics[angle=270,width=13.0cm]{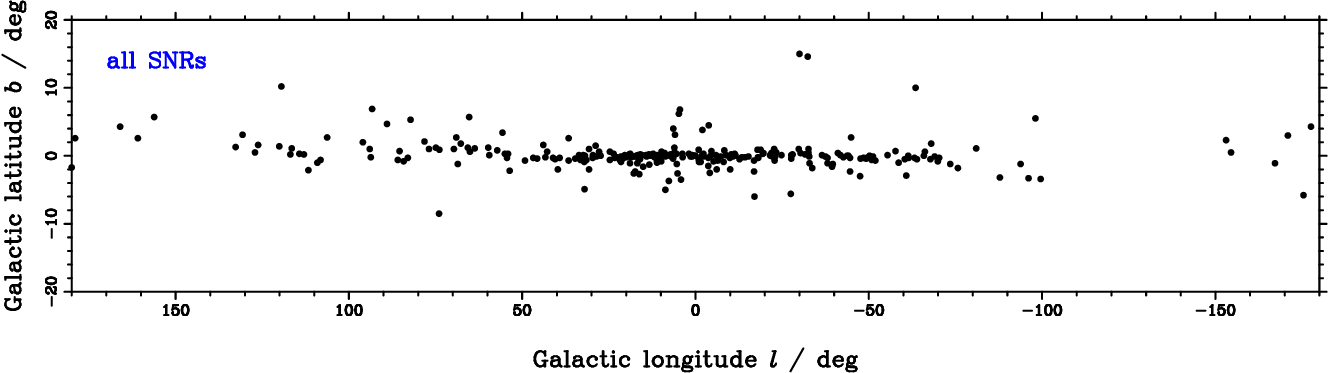}}
\bigskip
\centerline{\includegraphics[angle=270,width=13.0cm]{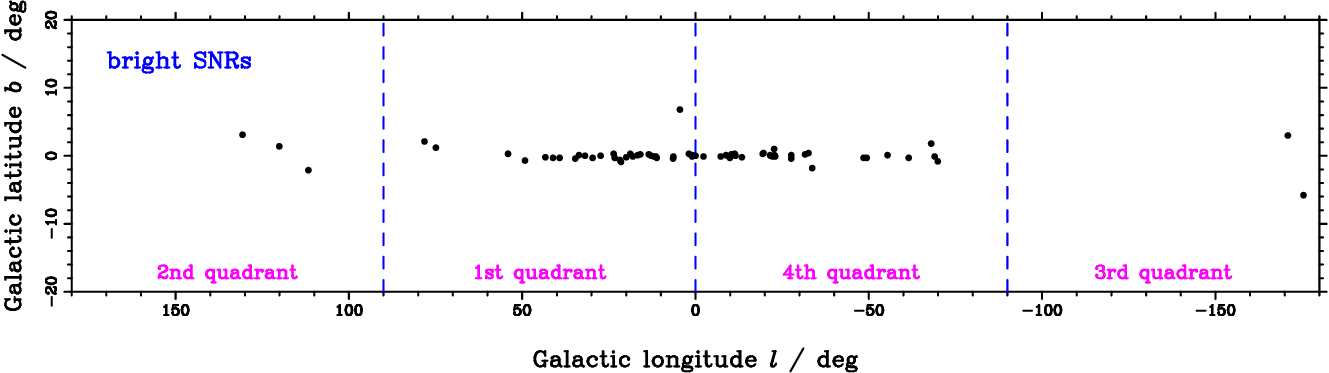}}
\bigskip
\caption{Galactic distribution of: (top) all 274 catalogued SNRs, and
(bottom) the brighter 68 remnants, with surface brightnesses above
$10^{-20}$ {\SigmaUnit} at 1~GHz. (Note that the $l$- and $b$-axes
are not on the same scale.)}\label{f:lb}
\end{figure}

\subsection{Selection effects}\label{s:selection}

Although some SNRs have first been identified at other than radio
wavelengths, in practice the vast majority have been identified from
radio observations (which, unlike the optical or X-rays, are not
affected by absorption). Furthermore, it is large-area radio surveys
that define the completeness of current SNR catalogues, not other
(better) observations, which cover specific targets, or are of only
limited areas of the Galactic plane.

For much of the Galactic plane -- $358^\circ < l < 240^\circ$, $|b| <
5^\circ$ -- the deepest, large-scale survey is that made at 2.7-GHz with
the Effelsberg 100-m telescope (\citealt{1990A&AS...85..633R,
1990A&AS...85..691F}). The rest of the Galactic plane has been covered
by a survey at 843~MHz made with MOST. Both these surveys identified
many new SNRs, see \citet{1988srim.conf..293R} and
\citet{1996A&AS..118..329W} respectively. New remnants identified from
these surveys were added to the 1991 and 1996 versions of my SNR
catalogue. For a SNR to be identified it needs to be bright enough to be
distinguished from the Galactic background. The approximate surface
brightness limit for the Effelsberg 2.7-GHz survey is thought to be
about $10^{-20}$ {\SigmaUnit} at 1~GHz.

Since 1991, when the new SNRs identified in the Effelsberg survey were
added to the catalogue, an additional 61 SNRs have been identified in
the region covered by this survey. Of these only 5 are brighter than a
surface brightness of $10^{-20}$ {\SigmaUnit} at 1~GHz (G0.3$+$0.0,
G1.0$-$0.1, G6.5$-$0.4, G12.8$+$0.0 and G18.1$-$0.1;
see \citealt{1994MNRAS.270..835G, 1996MNRAS.283L..51K,
2000ApJ...540..842Y, 2005ApJ...629L.105B, 2006ApJ...639L..25B}), with
the brightest being $\sim 3 \times 10^{-20}$ {\SigmaUnit}. As shown in
Figure~\ref{f:sigma}, the vast majority of the more recently identified
SNRs in the Effelsberg survey region are fainter than $10^{-20}$
{\SigmaUnit} at 1~GHz. The numbers of catalogued remnants with a surface
brightness above $10^{-20}$ {\SigmaUnit} at 1~GHz in the 1st and 4th
Galactic quadrants are 35 and 29 respectively, which are consistent
within Poisson statistics. Thus I take a surface brightness of
$10^{-20}$ {\SigmaUnit} at 1~GHz to be the approximate effective
$\itSigma$-limit of the current Galactic SNR catalogue.
Figure~\ref{f:lb} shows the observed distribution in Galactic
coordinates of both (a) all catalogued SNRs, and (b) the
68\footnote{Note that in \citet{2012AIPC.1505....5G}, there was an error
in the surface brightness of one SNR, so that 69 remnants were above
this nominal surface brightness limit to provide a sample of `bright'
SNRs. In fact there are 68 above this limit in the 2009 SNR catalogue.
This difference does not change the conclusions in
\cite{2012AIPC.1505....5G} significantly.} SNRs brighter than the
nominal surface brightness completeness limit of $10^{-20}$ {\SigmaUnit}
at 1~GHz. This clearly shows that taking the surface brightness
selection into account -- i.e.\ considering the brighter remnants only
-- the distribution of SNRs is more closely correlated towards both
$b=0^\circ$ and the Galactic Centre than might be thought if all SNRs
were considered. This is not surprising, as the lower radio emission
from the Galaxy in the 2nd and 3rd quadrants, and away from $b=0^\circ$,
means it is easier to identify faint SNRs in these regions. It is most
difficult to identify SNRs close to this nominal surface brightness
limit in regions of the Galactic plane with bright and complex
background radio emission, i.e.\ close to the Galactic Centre.

\begin{figure}
\centerline{\includegraphics[angle=270,width=10.0cm]{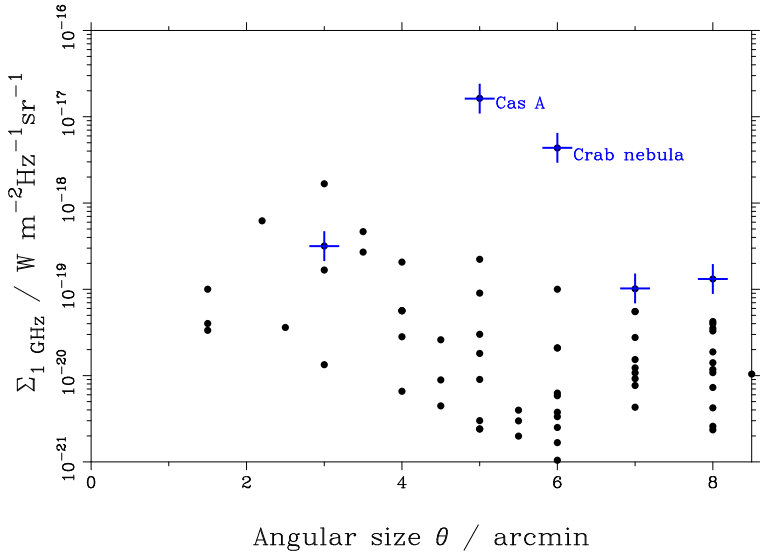}}
\bigskip
\caption{Surface brightness versus (mean) angular diameter for the
smaller catalogued SNRs. The remnants of the known `historical'
supernovae of AD 1054 (the Crab nebulae) 1181, 1572, 1604 plus Cas~A are
indicated by crosses. (The remnant of the supernova of AD 1006 is not
included, as it has a diameter of $\approx 30$~arcmin.)}\label{f:theta}
\end{figure}

Additionally, there is a selection effect that means that some small
angular size SNRs are overlooked. It is generally necessary to resolve a
SNR in order to recognise it structure, but not all of the Galactic
plane has been observed with sufficiently high resolution to resolve the
structure of all sources. The Effelsberg 2.7-GHz survey has a resolution
of 4.3~arcmin, making it difficult to recognised the structure of a
remnant unless it is $\sim 10$~arcmin or larger in extent. This means
that there is a deficit of small angular size SNRs, which is illustrated
by Figure~\ref{f:theta}. This shows the surface-brightness versus
angular diameter for the smaller SNRs in the current Galactic SNR
catalogue. The remnants of `historical' supernovae chronicled in the
last thousand years or so are indicated. All these remnants are
relatively close-by, as otherwise their parent supernova would not have
been seen visibly, and therefore they sample only a small part of the
Galactic disk. If these known young remnants were further away, they
would have the same surface brightness, but would be smaller in angular
size. Although there are some such remnants currently known -- e.g.\ the
very young SNR G1.9+0.3 (see \citealt{2008MNRAS.387L..54G,
2008ApJ...680L..41R, 2011ApJ...737L..22C}) -- there are fewer than
expected (see further discussion in \citealt{2005MmSAI..76..534G}).
Hence there is a selection effect against the identification of young
but distant SNRs in the Galaxy. Note that most of these missing young
remnants will be on the far side of the Galaxy, and therefore appear
nearer $b=0^\circ$ and to $l=0^\circ$. This is the region of the
Galactic plane where the background is brightest, and where there is
also more likely to be confusion with other Galactic sources along the
line of sight.

Of the 5 sources brighter than $10^{-20}$ {\SigmaUnit} -- i.e.\ above
the nominal surface brightness limit of the current SNR catalogue --
which have been identified since 1991 in the Effelsberg survey area, all
are close (within $20^\circ$) to the Galactic Centre. Moreover, 3 of
them are small, $\lesssim 8$~arcmin in diameter. Thus, it is likely that
the sample of 68 `bright' SNRs may be somewhat incomplete near the
Galactic Centre, due to (i) missing young but distant remnants, and (ii)
the difficulty of identifying remnants near the surface brightness limit
in this region of the Galaxy, with a relativity bright and complex
background.

The 2009 version of the catalogue includes remnants identified in the
refereed literature published up to the end of 2008. Since then some
other remnants have been identified (e.g.\ G25.1$-$2.3 and G178.2$-$4.2,
\citealt{2011A&A...532A.144G}; G35.6$-$0.4,
\citealt{2009MNRAS.399..177G}; G64.5+0.9, \citealt{2009MNRAS.398..249H};
G296.7$-$0.9, \citealt{2012MNRAS.419.2623R}; G308.3$-$1.4,
\citealt{2012ApJ...750....7H}; G310.5$+$0.8,
\citealt{2011Ap&SS.332..241S}), but none of these are clearly brighter
than $10^{-20}$ {\SigmaUnit} at 1~GHz.

\subsection{The `$\itSigma{-}D$' relation}\label{s:sigmad}

To directly construct the Galactic distribution of SNRs it is necessary
to know the distance to each remnant. Distances are only available for
about 20\% of currently known SNRs, and so the surface
brightness--linear diameter -- or `$\itSigma{-}D$' -- relation has often
been used instead. This provides an estimated linear size for a remnant
from its \emph{observed} surface brightness, using the $\itSigma{-}D$
correlation seen for SNRs with known distances. This correlation is
usually parameterised as
$$
  \itSigma = C D^{-n}
$$
as physically small SNRs tend to have larger surface brightnesses than
larger ones. As is discussed in \citet{2005MmSAI..76..534G}, much of
this correlation is arguably due to a $D^{-2}$ bias due to the fact that
$\itSigma \propto L/D^2$, where $L$ is the luminosity of the remnant. In
practice, however, there are several issues with the `$\itSigma{-}D$'
relation. First, SNRs show a wide range of physical diameters for a
given surface brightness, approximately an order of magnitude in
range.
This means that a distance derived for an individual remnant is quite
inaccurate. Second, due to the observational selection effect discussed
above, the range of properties of SNRs may be larger than is evident
from currently identified remnants, as small angular size, faint
remnants are particularly difficult to identify. Third, as has been
discussed in \citet{2005MmSAI..76..534G}, some `$\itSigma{-}D$' studies
have used inappropriate least-square straight line regressions. As there
is a larger scatter in the $\itSigma{-}D$ plane, regressions minimising
deviations in $\itSigma$ give quite a different correlation than one
minimising deviations in $D$ (e.g. see \citealt{1990ApJ...364..104I}).
Since the $\itSigma{-}D$ relation is used to predict a value for $D$ from
the $\itSigma$ value for an individual remnant, then minimising deviations
in $D$ should be used. \citet{1998ApJ...504..761C} minimised the
deviations in $\itSigma$, and obtained a $\itSigma{-}D$ relation with $n =
2.64 \pm 0.26$ (for 37 `shell' remnants, including Cas~A), whereas a
significantly steeper relation with $n = 3.53 \pm 0.33$ is obtained if
deviations in $D$ are used. This means that fainter remnants -- which
are the majority, see Figure~\ref{f:sigma} -- have their diameters, and
hence distances, \emph{overestimated} if a $\itSigma{-}D$ relation
minimising deviations in $\itSigma$ is used.

\begin{figure}
\centerline{\raisebox{-\height}{(a)}
 \includegraphics[bb=96 61 406 783,angle=270,width=10.5cm,clip=]{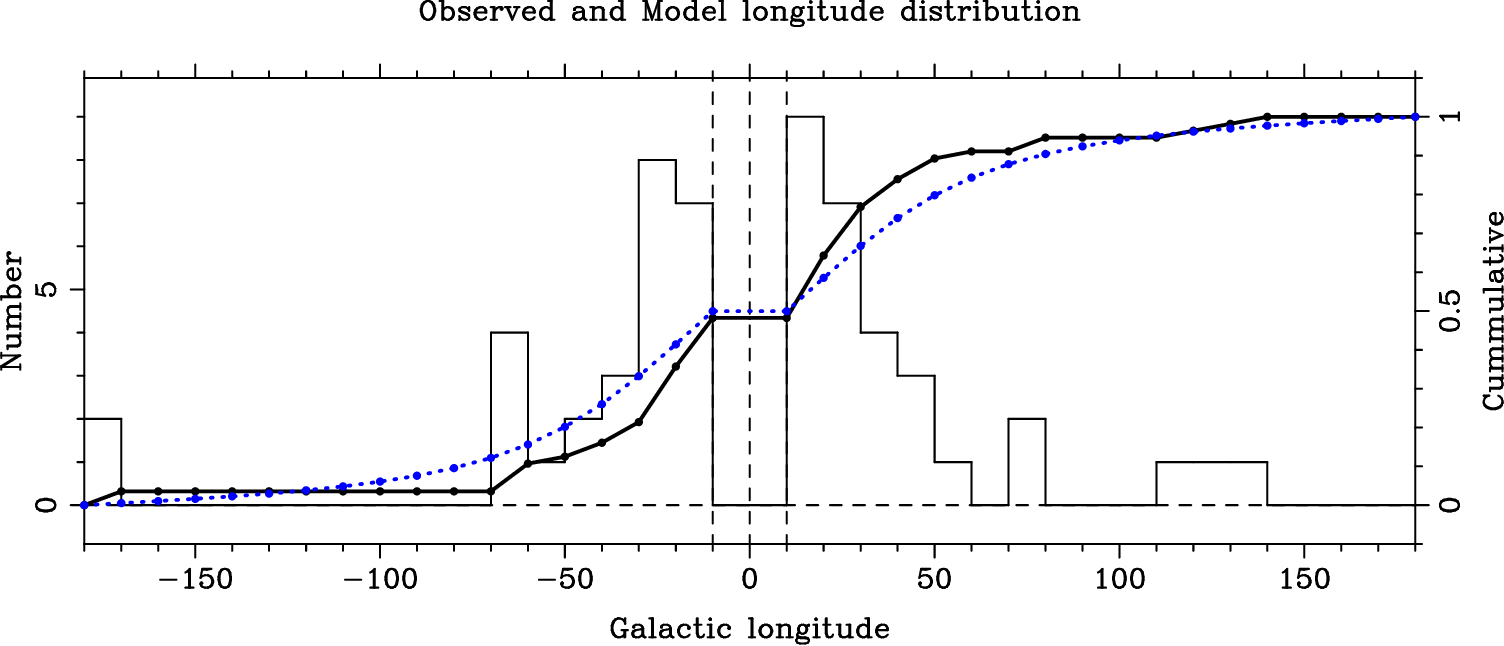}}
\bigskip
\centerline{\raisebox{-\height}{(b)}
 \includegraphics[bb=115 61 406 783,angle=270,width=10.5cm,clip=]{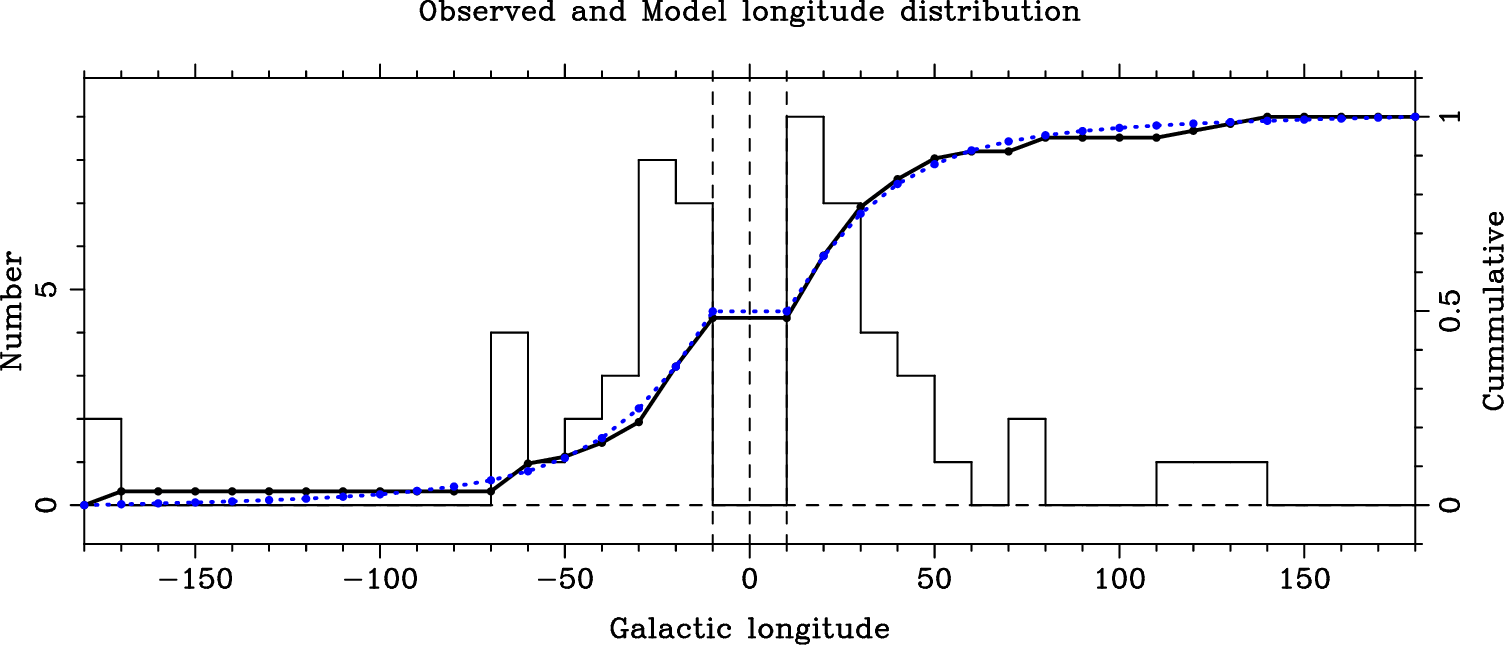}}
\bigskip
\centerline{\raisebox{-\height}{(c)}
 \includegraphics[bb=115 61 406 783,angle=270,width=10.5cm,clip=]{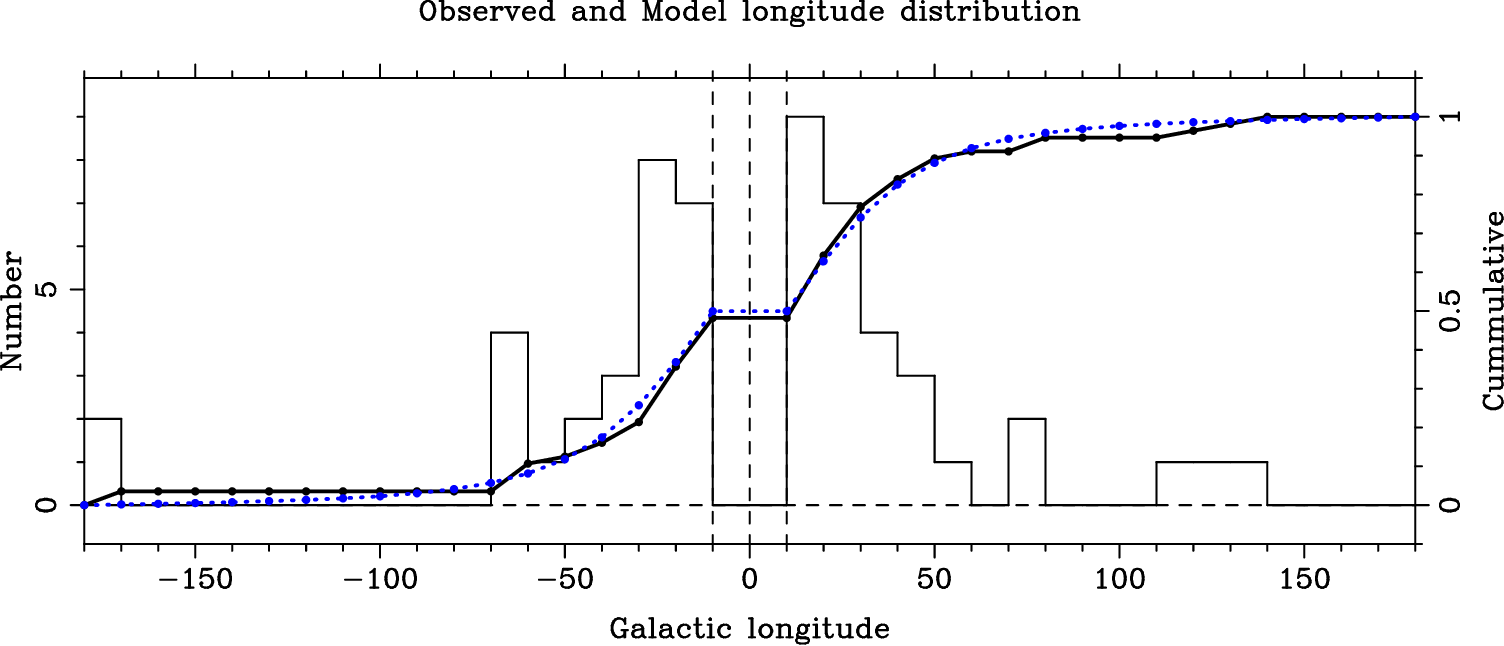}}
\bigskip
\caption{The $l$-distribution of the 56 `bright' Galactic SNRs --
excluding those with $|l| \le 10^\circ$ -- shown as (i) histogram (left
scale), and (ii) cumulative fraction, solid line (right scale). In
addition the cumulative fraction for a model distribution is also
plotted, dotted line (right scale). The three models presented are for
the surface density of SNRs varying with Galactocentric radius, $R$, as
(a) $\propto ({R}/{R_\odot})^{2.0} \exp \left[ -3.5(R-R_\odot)/R_\odot
\right]$ (as derived by \citealt{1998ApJ...504..761C}), (b) $\propto
({R}/{R_\odot})^{0.7} \exp \left[ -3.5(R-R_\odot)/R_\odot \right]$, and
(c) $\propto ({R}/{R_\odot})^{2.0} \exp \left[ -5.1(R-R_\odot)/R_\odot
\right]$.}\label{f:sd}
\end{figure}

\section{The Galactic distribution of SNRs}\label{s:gd}

The direct approach to deriving the distribution of Galactic SNRs is to
use the $\itSigma{-}D$ relation to derive distances to individual
remnants, and then construct the 3-D distribution of remnants. However,
because of the large range of diameters shown for remnants with similar
surface brightnesses, the $\itSigma{-}D$ relation does not provide
reliable distances to individual remnants. Moreover there are the
observational selection effects discussed in Section~\ref{s:selection},
which mean that it is not possible to use treat catalogued remnants with
equal weight. Instead, the approach I use is to consider only brighter
remnants above the nominal surface-brightness limit, and compare their
distribution in Galactic longitude with that expected from various
models. This approach does not need distance estimates for individual
SNRs. Because of the possible remaining selection effects close to the
Galactic centre, the region $|l| \le 10^\circ$ is excluded from the
analysis presented here, leaving 56 brighter remnants.

\begin{figure}
\centerline{\includegraphics[angle=270,width=11.0cm]{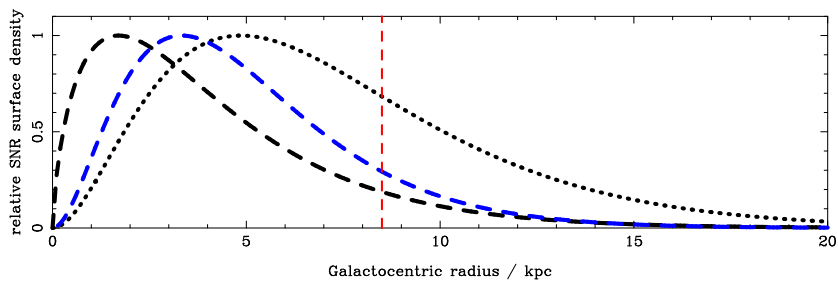}}
\bigskip
\caption{The distribution (in terms of surface density) of SNRs with
Galactocentric radius, $R$, for the three power-law/exponential models
shown in Figure~\ref{f:sd} and discussed in Section~\ref{s:gd}: dotted
line for \citet{1998ApJ...504..761C}'s distribution (a), and dashed
lines for models (b) and (c).}\label{f:radial}
\end{figure}

One model from the distribution of SNRs (and other star formation
tracers, e.g.\ pulsars and star formation regions, see
\citealt{2000A&A...358..521B, 2006MNRAS.372..777L}) is a two parameter
power-law/exponential radial distribution for the density of SNRs with
Galactocentric radius, $R$, of the form
$$
  \propto \left( \frac{R}{R_\odot} \right)^A
  \exp \left[ -B \frac{(R-R_\odot)}{R_\odot} \right]
$$
(with $R_\odot = 8.5$~kpc, the distance to the Galactic Centre).
The observed distribution in $l$ of SNRs with $\itSigma > 10^{-20}$
{\SigmaUnit} from \citet{2009BASI...37...45G} excluding those with $|l|
\le 10^\circ$ is shown in Figure~\ref{f:sd}, along with the
distributions from three different power-law/exponential radial models.
The models are (a) $A=2.0$, $B=3.5$ (i.e.\ the distribution obtained by
\citealt{1998ApJ...504..761C}), (b) $=0.7$, $B=3.5$ (i.e.\ the same
value for $B$ as in (a), but adjusting $A$ for a best least square fit
between the observed and cumulative distributions), and (c) $=2.0$,
$B=5.1$ (i.e.\ the same value for $A$ as in (a), but adjusting $B$ for a
best fit). From Figure~\ref{f:sd}(a) it is clear that the
power-law/exponential distribution obtained by
\citet{1998ApJ...504..761C}, is broader than the observed distribution
of `bright' SNRs above the nominal surface brightness limit of current
SNR catalogues (which is to be expected, given the systematic effect due
to the regression used by \citealt{1998ApJ...504..761C} noted in
Section~\ref{s:sigmad}). Models (b) and (c) have very similar least
squares differences from the observed cumulative distribution, but
correspond to somewhat different distributions in Galactocentric radius.
Figure~\ref{f:radial} shows the distribution of Galactic SNRs
against Galactocentric radius for the three models. This shows that
there is degeneracy between the parameters $A$ and $B$ in the
power-law/exponential distribution model.

\section{Conclusions}

The lack of distances to most known Galactic SNRs, plus observational
selection effects, means that it is difficult to derive the distribution
of SNRs in our Galaxy directly. However, by considering `bright' SNRs --
i.e.\ those not strongly affected by selection effects -- constraints on
the Galactic distribution of SNRs can be obtained, by comparison of
their $l$-distribution with that expected from models. This shows that
the Galactic distribution of SNRs obtained by
\citet{1998ApJ...504..761C} is too broad.

\begin{acknowledgments}
I thank Irina Stefan for useful discussions.
\end{acknowledgments}


\label{lastpage}

\end{document}